\documentclass[]{aa}

\usepackage{rotating}
\usepackage{graphicx}
\usepackage{sidecap}
\usepackage{subcaption}
\usepackage{url}
\usepackage[colorlinks=true]{hyperref}
\usepackage{soul}
\hypersetup{allcolors=blue}
\usepackage{float}
\usepackage[tracking=true,kerning=true,shrink=50]{microtype}

\usepackage{lipsum}
\usepackage{txfonts}
\usepackage[]{xcolor}
\usepackage[export]{adjustbox} 

\bibpunct{(}{)}{;}{a}{}{,}
\DeclareUnicodeCharacter{00A0}{~}
\usepackage[us,12hr]{datetime}
\usepackage{ulem}

\usepackage{cancel}
\usepackage{color}

\definecolor{deepmagenta}{rgb}{0.8, 0.0, 0.8}
\definecolor{green}{rgb}{0.1, 0.7, 0.1}

\usepackage{algorithm}
\usepackage{algpseudocode}

\begin{document}

    \title{Designing wavelength sampling for Fabry-Pérot observations}
    \subtitle{Information-based spectral sampling}

   \author{C. J. D\'iaz Baso
          \inst{1,2,3}
          \and
          L. Rouppe van der Voort
          \inst{1,2}
          \and
          J. de la Cruz Rodr\'iguez
          \inst{3}
          \and
          J. Leenaarts
          \inst{3}
          }
   \institute{
   Institute of Theoretical Astrophysics,
   University of Oslo, %
   P.O. Box 1029 Blindern, N-0315 Oslo, Norway
   \and
   Rosseland Centre for Solar Physics,
   University of Oslo, %
   P.O. Box 1029 Blindern, N-0315 Oslo, Norway
   \and
   Institute for Solar Physics, Dept. of Astronomy, Stockholm University, AlbaNova University Centre, SE-10691 Stockholm, Sweden
   \\
   \email{carlos.diaz@astro.uio.no}
   }

   \date{Draft: compiled on \today\ at \currenttime~UT}

   \authorrunning{D\'iaz Baso et al.}

  \abstract
   {Fabry-Pérot interferometers (FPIs) have become very popular in solar observations because they offer a balance between cadence, spatial resolution, and spectral resolution through a careful design of the spectral sampling scheme according to the observational requirements of a given target. However, an efficient balance requires knowledge of the expected target conditions, the properties of the chosen spectral line, and the instrumental characteristics.}
   {Our aim is to find a method that allows finding the optimal spectral sampling of FPI observations in a given spectral region. The selected line positions must maximize the information content in the observation with a minimal number of points.}
   {In this study, we propose a technique based on a sequential selection approach where a neural network is used to predict the spectrum (or physical quantities, if the model is known) from the information at a few points. Only those points that contain relevant information and improve the model prediction are included in the sampling scheme.}
   {We have quantified the performance of the new sampling schemes by showing the lower errors in the model parameter reconstructions. The method adapts the separation of the points according to the spectral resolution of the instrument, the typical broadening of the spectral shape, and the typical Doppler velocities. The experiments using the \ion{Ca}{ii} 8542\,\AA\ line show that the resulting wavelength scheme naturally places (almost a factor 4) more points in the core than in the wings, consistent with the sensitivity of the spectral line at each wavelength interval. As a result, observations focused on magnetic field analysis should prioritize a denser grid near the core, while those focused on thermodynamic properties benefit from a larger coverage. The method can also be used as an accurate interpolator, to improve the inference of the magnetic field when using the weak-field approximation.}
   {Overall, this method offers an objective approach for designing new instrumentation or observing proposals with customized configurations for specific targets. This is particularly relevant when studying highly dynamic events in the solar atmosphere with a cadence that preserves spectral coherence without sacrificing much information.}


   \maketitle


\section{Introduction}\label{sec:intro}

Solar processes manifest themselves on a wide range of spatial, temporal, and energetic scales, and high-quality measurements are crucial for understanding the nature of these phenomena. The magnetic field plays a key role in transporting mass and energy to the upper layers. However, establishing a precise quantification of the impact of magnetic fields is a challenge since that needs spectropolarimetric measurements with high spatial (sub arcsec), spectral ($<100$\,m\AA), and temporal ($<10$\,s) resolution along with high polarimetric sensitivity ($<0.1$\% of the intensity) \citep{Jaime2017, Iglesias2019}.

This study is motivated by the instrumental requirements arising from solar observations and the trade-offs associated mainly with the high dimensionality of the data (image space, spectral information, polarimetry, and time evolution), signal-to-noise ratio (SNR), and resolution limitations (see \citealt{Iglesias2019} for a detailed review). Historically, the most commonly used instrumentation for spectral mapping in optical solar spectropolarimetry are grating and filter spectrograph systems. Grating spectrographs reduce a spatial dimension by using a long slit, where the entire spectrum is captured. However, they have to scan the solar surface to generate a bi-dimensional map of an extended source. On the other hand, filtergraphs can be used for narrow-band imaging, presently the most popular being Fabry-Pérot interferometers (FPIs). The core of FPIs are highly reflective cavities with a known tunable thickness that allows only one wavelength position to be observed at a time. By scanning the spectral range, it is possible to obtain bi-dimensional maps of the field of view. This is a limiting factor since all acquisitions within a line scan should be as simultaneous as possible. Therefore, observing a dynamic and complex region such as the chromosphere involves a problematic trade-off between wavelength coverage and acquisition time. This trade-off has led to the design of specific observing schemes depending on the solar target under study \citep[see, e.g.,][]{delaCruz2012A&A, Felipe2019A&A}.

In general, FPIs have become very popular because a balanced trade-off between cadence, spatial resolution, and spectral resolution can be found by designing a spectral sampling scheme as a function of the target. This choice is particularly important when observing flares, umbral-flashes, or similar fast-evolving events and preserving spectral coherence \citep{Felipe2018A&A, Kuridze2018ApJ, Yadav2021, Schlichenmaier2022}. At ground-based telescopes, examples of this type of instrument are the CRisp Imaging SpectroPolarimeter \citep[CRISP;][]{Scharmer2008} and the CHROMospheric Imaging Spectrometer \citep[CHROMIS;][]{Scharmer2017} at the Swedish 1-meter Solar Telescope \citep[SST;][]{Scharmer2003}, the BIdimensional Interferometric Spectrometer \citep[IBIS;][]{Cavallini2006} at the Richard B. Dunn Solar Telescope (DST) and the GREGOR Fabry-Pérot Interferometer \citep[GFPI;][]{Puschmann2012}, but also adopted by the design team of space telescopes such as the Polarimetric and Helioseismic Imager on the Solar Orbiter mission \citep[PHI; ][]{Solanki2020A&A}, and the Tunable Magnetograph on the Sunrise-III mission \citep[TuMag; ][]{Magdaleno2022}. Furthermore, Fabry-Pérot systems remain crucial instruments for providing large FOV observations to the next generation 4-m class of telescopes such as the Daniel K. Inouye Solar Telescope \citep[DKIST;][]{DKIST2020} {with the Visible Tunable Filter \citep[VTF;][]{Schmidt2014}} and the European Solar Telescope \citep[EST;][]{QuinteroNoda2022} with its Tunable Imaging Spectropolarimeter (TIS).

The efficient design of an observing scheme that includes the minimum number of measurements allowing fast spectral line scanning but also accurate estimation of the physical parameters that give rise to the spectra requires knowledge of the expected target conditions, the properties of the chosen spectral line, and the instrumental characteristics. This task is exactly the aim of feature selection, which is a subfield of statistical machine learning in which the main goal is to maximize the accuracy in estimating some objective of interest using the smallest possible number of input features \citep{Guyon2003, 2016arXiv160107996L, Settles2012}. This strategy is advantageous in modern machine learning problems to improve the computational efficiency and to avoid input features that contain {irrelevant or redundant} information. There are several different strategies for deciding which features are most informative, and there already exist frameworks that include some of these solutions, e.g. MLextent \citep{raschkas_2018_mlxtend} and Scikit-learn \citep{scikit-learn}. This idea has been explored recently by \cite{Lim2021ApJ} and \cite{Salvatelli2022} to analyze how much redundant information is present in the channels of the Atmospheric Imaging Assembly \citep[AIA;][]{aia2012} of the Solar Dynamics Observatory \citep[SDO;][]{sdo2012} by predicting every filter-bands from the rest \citep{Panos2021b}.

In this study, we explore the design of spectral sampling schemes to choose the most informative points in the wavelength range. The paper is organized as follows. We begin with a brief introduction to the method and how we implement the approach on two examples of different complexity, a Gaussian absorption profile and the application on the \ion{Ca}{ii} 8542\,\AA\ line when a pre-existent dataset with high spectral resolution is provided (Sect.~\ref{sec:spectra}). The new sampling is compared to a uniformly spaced sampling in different ways. We also explore the idea of optimizing a sampling scheme for a particular physical parameter if the physical model is available in Sect.~\ref{sec:model}. Finally, we briefly discuss the implications of this work and describe possible extensions and improvements (Sect.~\ref{sec:conclusions}).

\section{Information-based spectral sampling}

\subsection{Using spectral information}\label{sec:spectra}

\subsubsection{Description of the method}

The main task of feature selection is the procedure of selecting the important features from the data without degrading the performance of our task. In our case, we aim to sample the spectral line by scanning parts containing different information and avoid repeating areas with similar information (e.g., nearby regions in wavelength). In this way, we will be able to scan a spectral line with the minimum set of points that will allow us a fast spectral line scan while keeping as much information as possible. {This approach known as unsupervised selection, as it does not require prior knowledge of how the data will be used, but rather focuses on extracting information from pre-existing datasets.}

Traditionally, the task of sampling a functional form efficiently has been used in cases where no other information than the assumption of continuity and smoothness is given a-priori. This would mean that we do not have previous observations or simulated data that could indicate the usual shapes of the spectral line and take advantage of them. So these methods are commonly based on the uncertainty of predictor models trained on the very few examples \citep{LewisC94}, struggling to provide a useful or efficient solution. Fortunately, today we have a growing network of telescopes as well as a large number of radiative transfer and simulation codes \citep[e.g.,][]{Uitenbroek2001, Vogler2005A&A, Gudiksen2011A&A, Stepan2013A&A, Leenaarts2009ASPC, 2015A&A...577A...7S, Millic2018A&A,2020A&A...638A..79N,2021ApJ...917...14O,2021ApJ...911...71A, 2022A&A...664A..91P} that can provide spectra of our line of interest and incorporate this information in the process of designing the sampling scheme. 

There are several ways to tackle this problem, but we would like to use a computationally feasible procedure to avoid an exhaustive search of all the possible combinations, even though the final configuration may be slightly suboptimal. The {simplest approach} is probably the sequential forward selection \citep[SFS;][]{Whitney1971, Pudil1994PaReL} where the algorithm adds iteratively the most informative point to the list of previous points.
The most valuable point to add will be the one our model is not able to predict correctly, i.e., the point where the error is maximum. Then, this point is included in the output list, and the remaining points are considered given this new information. This is repeated until a pre-specified number of points has been included. In practice, we aim to estimate an optimal sampling scheme for a spectral line that will have different shapes, so we will evaluate the average error. Although our criterion is the average error because of its {simplicity in the implementation}, other criteria can also be used, such as the Mutual Information \citep{Panos2021a}, which measures the correlation between different variables.

Once the motivation has been explained, we can describe the algorithm in more detail. Let $L=\{x_m,\{y_m^k\}_{k=1}^K\}_{m=1}^{M}$ represent our data consisting of $M$ spectral points in a dataset of $K$ examples of different line profiles. 
{ Ideally the wavelength spacing of the dataset should be as fine as possible given that the optimized scheme will lay on this grid. In practice, an equidistant spacing better than two or four wavelength points per spectral resolution element is a good starting point.}
{ When considering the number of $K$ examples in a dataset, it is important to prioritize the diversity of the examples over the total number of examples. A diverse dataset that includes all the relevant features and characteristics we expect to encounter in real-world observations is essential for developing a versatile scheme.}
Each $x_m$ represents the wavelength information, and $y_m$ contains the information of the observables (which can be just intensity or all the Stokes parameters). Let $U$ be a subset of $L$ which will include the wavelength points of our final scheme. In the beginning, $U_0$ can be empty or contain, for example, the values at the edges of the interval. A non-linear model predictor, which we chose to be a neural network $f_\theta$ with internal parameters $\theta$, will be trained every time a new point is required. The input of the predictor is the subset $U_0$ with the observables at a few wavelength points at this iteration, and the output should be the observables at all the $M$ spectral points. The wavelength location where the mean squared error (mse) between the full dataset $L$ and the prediction $f_\theta(U_0)$ of the network is largest is chosen to be the next instance to add to $U_1$, i.e. argmax$(\sigma_{0}^2)$. Then the same procedure is repeated until the total number of points is reached. The total number of points will determine the temporal cadence of the observation, so it is a number that should be chosen according to the requirements of the target. Thus, the pseudocode for finding an optimal sampling scheme for a given dataset would be as follows:
\begin{algorithm}
	\caption{Current error-based spectral sampling} 
	\begin{algorithmic}[1]
	\State $L$ = pool of available points
	\State $U_0$ = set of initial points 
	\State $P$ = number of points to add to the scheme
		\For {$p =1,2,\ldots$ $P$}
			\State train the predictor $f_\theta$ such that $L$ $\sim f_\theta$($U_{p-1}$) 
			\State evaluate the criteria $\sigma_{p-1}^2$ = mse($L$, $f_\theta$($U_{p-1}$)\ )
			\State identify the most informative point: $i$ = argmax$(\sigma_{p-1}^2)$
			\State add the new wavelength point $U_{p}=U_{p-1}+\{x_i,\{y_i^k\}_{k=1}^K\}$ 
		\EndFor
	\end{algorithmic} 
\end{algorithm}

{This method is based on a forward selection, which is computationally more efficient than the alternative  backward elimination.} 
This algorithm can be accelerated, for example, by avoiding evaluating the criteria in those points we already have in the scheme. In the following example, we have decided to evaluate all the points so we can check that the predictor is really using the information in the sampling scheme. We should note that the total variance $\sigma_p^2$ includes the intrinsic noise of the data (which does not depend on the model), the flexibility of the model itself (which is invariant given a fixed model class), and the model variance, which is the remaining component that we try to minimize to improve the generalization of the model. We will see later how the other components begin to dominate as the number of points increases making the process much less efficient. Regarding the neural network architecture, in the following examples we have used $f_\theta$ as a simple residual network \citep[ResNet,][]{ResNet2015} with 2 residual blocks, and 64 neurons per layer and ELU as the activation function. We trained the models for {10k} epochs with the Adam optimizer \citep{Kingma2014} and a learning rate of {$5\cdot10^{-4}$}. Although several variants of this implementation exist (e.g., MLextent, Scikit-learn), the restricted flexibility of those frameworks do not allow us to perform the following experiments, so we have implemented them in PyTorch\footnote{Our implementation can be found in the following repository: \\ \url{https://github.com/cdiazbas/active_sampling}.}.

\subsubsection{Gaussian example}

To show a practical application of this method, we have created an artificial set of 20000 Gaussian-shaped profiles, with the amplitude drawn from a uniform distribution between 0.2 and 1 units, with the velocity following a normal distribution with 0.0 mean and a standard deviation of 5 units, and with the width from a uniform distribution between 1.5 and 2.5 units. We have assumed that our full-resolution database is sampled with 100 points over a range of 20 units.

\begin{figure}[htp!]
\centering
\includegraphics[width=\linewidth]{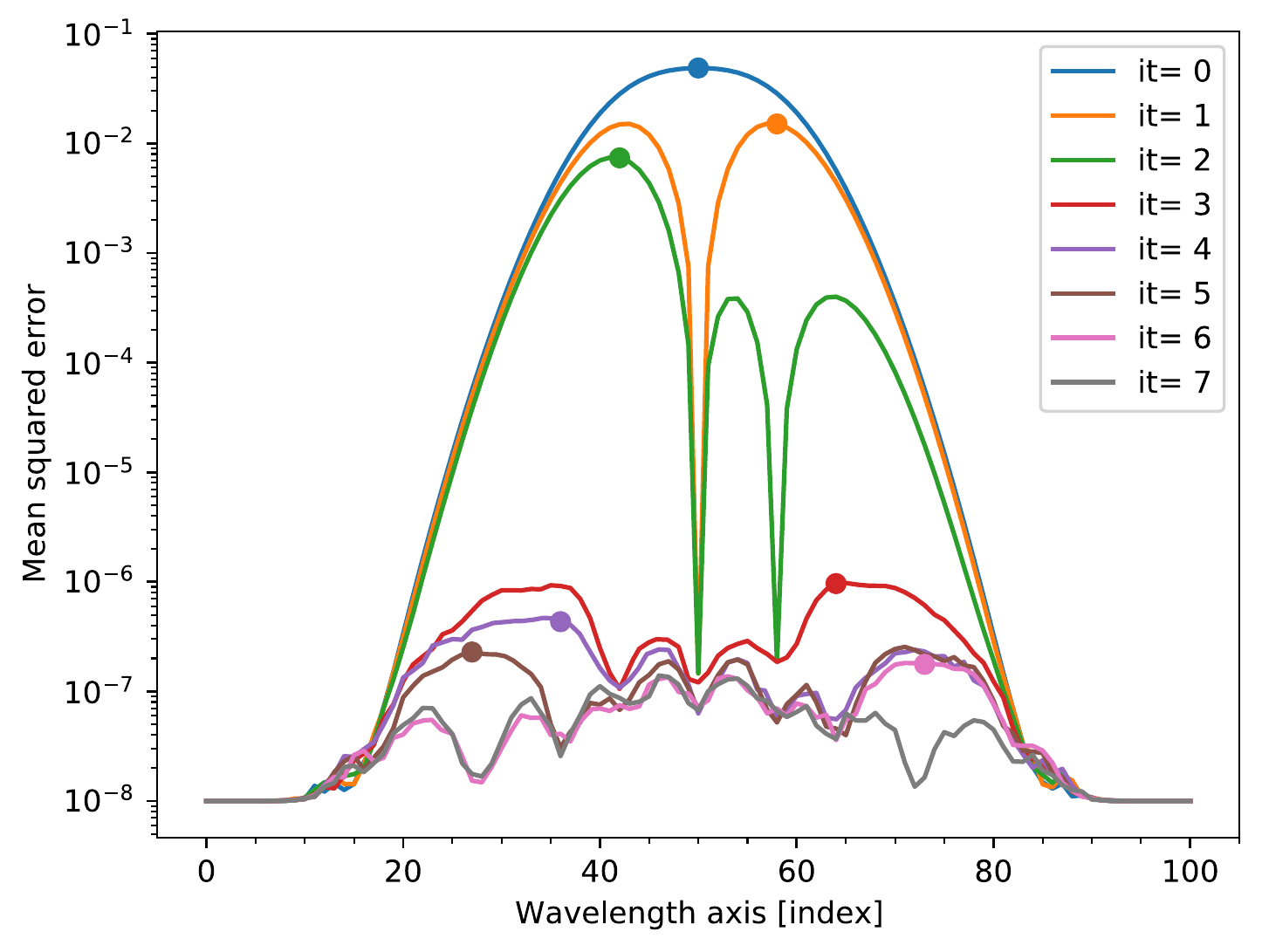}
\caption{Mean square error of the network used as a predictor of all points in the range. At each iteration, the location of the point at the maximum location is added to the list of sampling points.} \label{fig:gauss1}
\end{figure}

Then, we ran the method described before on this dataset iteratively for a total of 9 points, starting with the two points at the edges of the interval. This process can be seen in Fig.~\ref{fig:gauss1} by the mean square error at every iteration and the maximum indicated with a dot with the same color. In each iteration, the network prediction is improved after including the information at the location of the maximum error in the previous iteration. Given the simple shape of the Gaussian, only a few points are required to estimate the location of the profile and this is observed in the rapid decrease of the error after adding 3 points. As we add more points, the error over the entire range becomes more homogeneous, and the placement of the last points will depend on the accuracy of the network. 

\begin{figure}[htp!]
\centering
\includegraphics[width=0.9\linewidth]{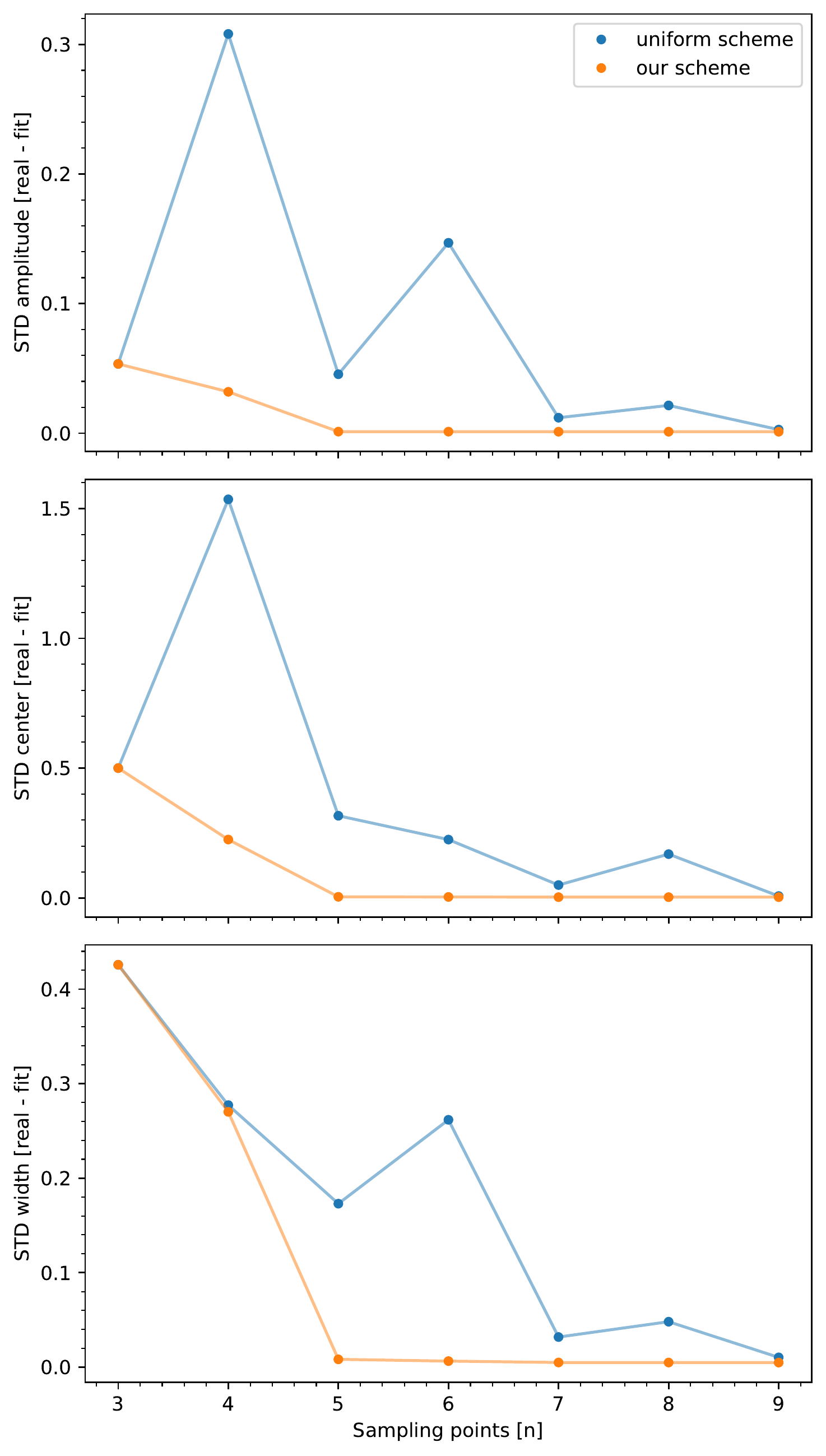}
\caption{Average error in the model parameters of the Gaussian fitting when using a uniform or an uncertainty-based sampling for a set of 10000 profiles.} \label{fig:gauss2}
\end{figure}

In the following, we verify the performance of the resulting sampling compared to a uniformly-spaced sampling, which is taken as our baseline. For this purpose, we fitted a Gaussian model using gradient-based minimization methods to the 10000 profiles in the same dataset using the two different sampling schemes. The average errors of the Gaussian parameters (here the amplitude, center, and width) are shown in Fig.~\ref{fig:gauss2}. This figure shows how the errors in the parameters using the information-based scheme are much smaller than those of the uniform sampling. For a low number of points, this difference is large and becomes smaller the more points we have in the range as expected. Given the sequential way of choosing points, the more points we add, the better it always gets. On the other hand, the uniform sampling only improves when the scheme has points in particular configurations. For example, odd numbers place a point at the center of the interval that coincides with the location of the average center of the profiles. After nine points, both sampling schemes reach a similar precision in this simplified example.

\begin{figure*}[htp!]
\includegraphics[width=0.97\linewidth,right,trim={0cm 1.2cm 0cm 0cm},clip]{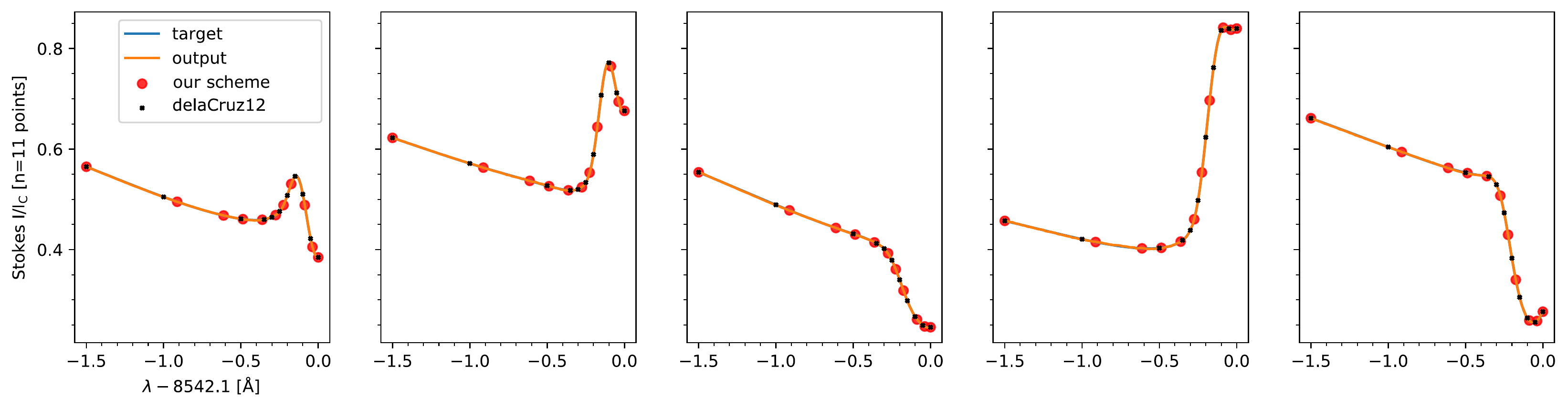}
\includegraphics[width=0.978\linewidth,right,trim={0cm 0.3cm 0cm 0cm},clip]{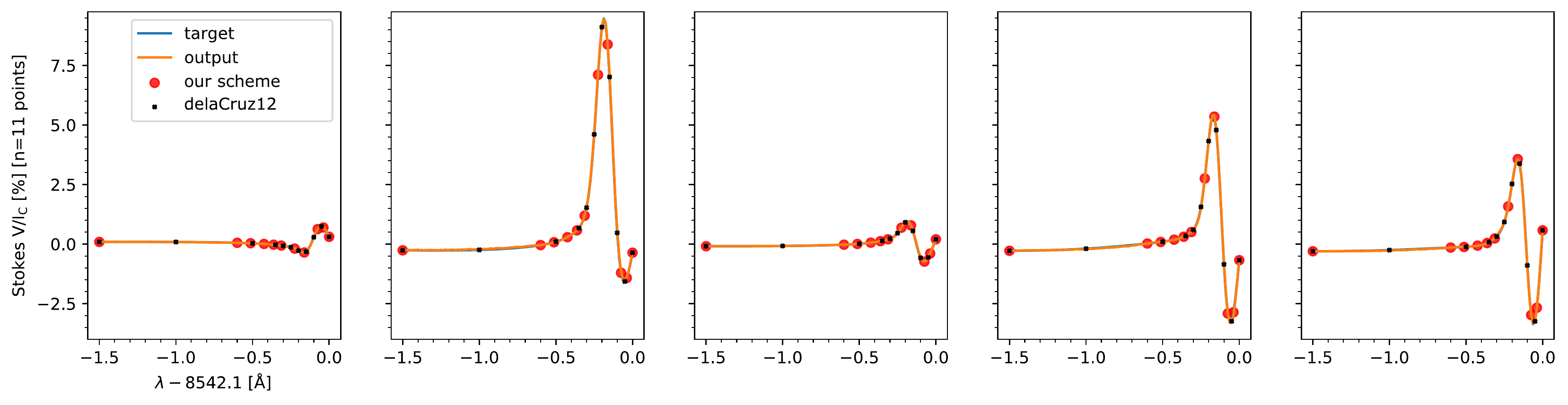}
\caption{Output sampling scheme as seen in five different profiles for Stokes $I$ (upper panel) and Stokes $V$ (lower panel) for the \ion{Ca}{ii} 8542\,\AA\ line. The sampling scheme suggested in \citet{delaCruz2012A&A} is also shown for comparison. The predicted profiles (output) are not visible because they are aligned behind synthetic profiles (target).} 
\label{fig:ca_sampling}
\end{figure*}

\begin{figure}[ht!]
\centering
\includegraphics[width=1.0\linewidth,trim={0cm 1.2cm 0cm 0cm},clip]{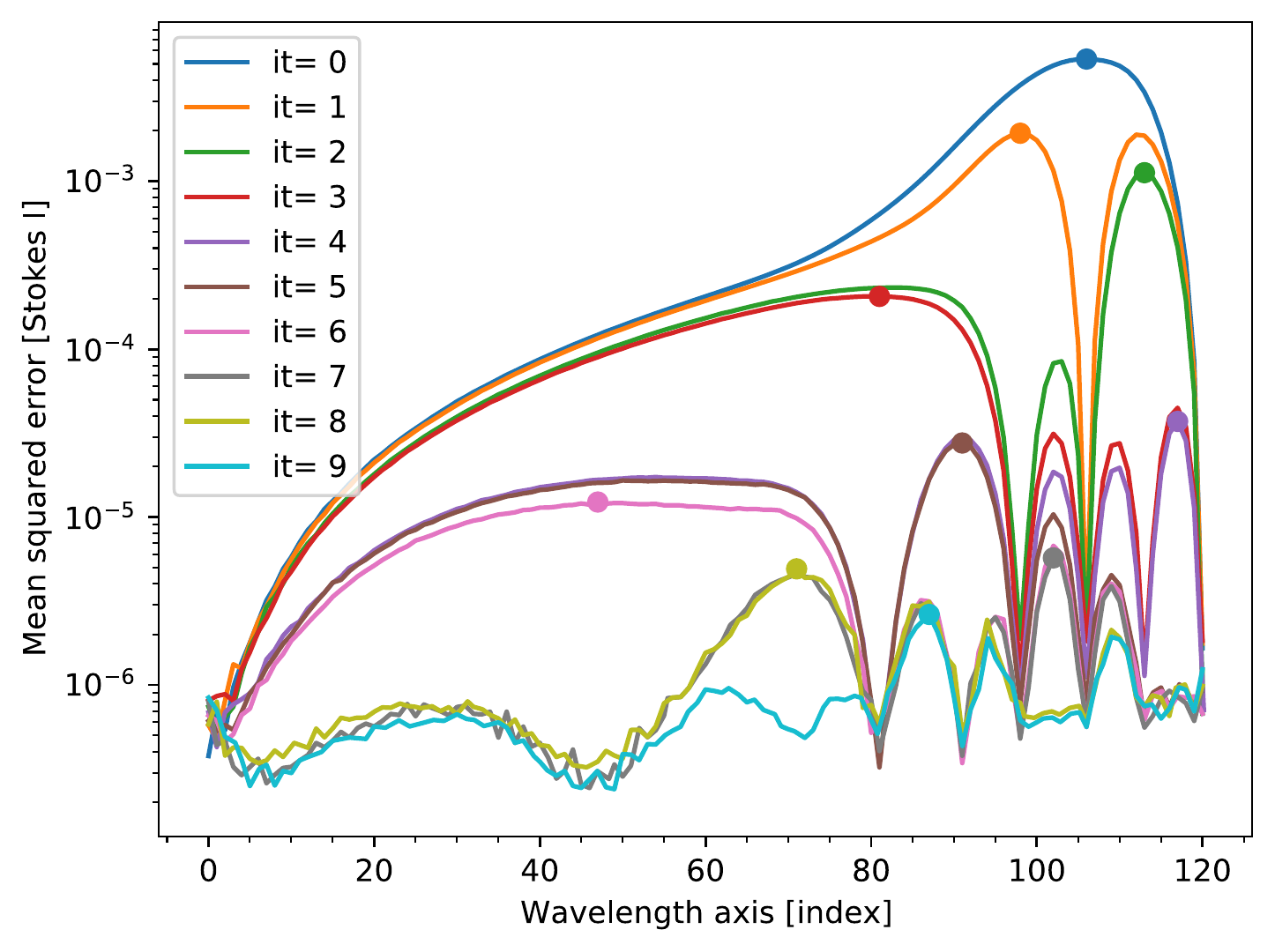}
\includegraphics[width=1.0\linewidth,trim={0cm 0.3cm 0cm 0cm},clip]{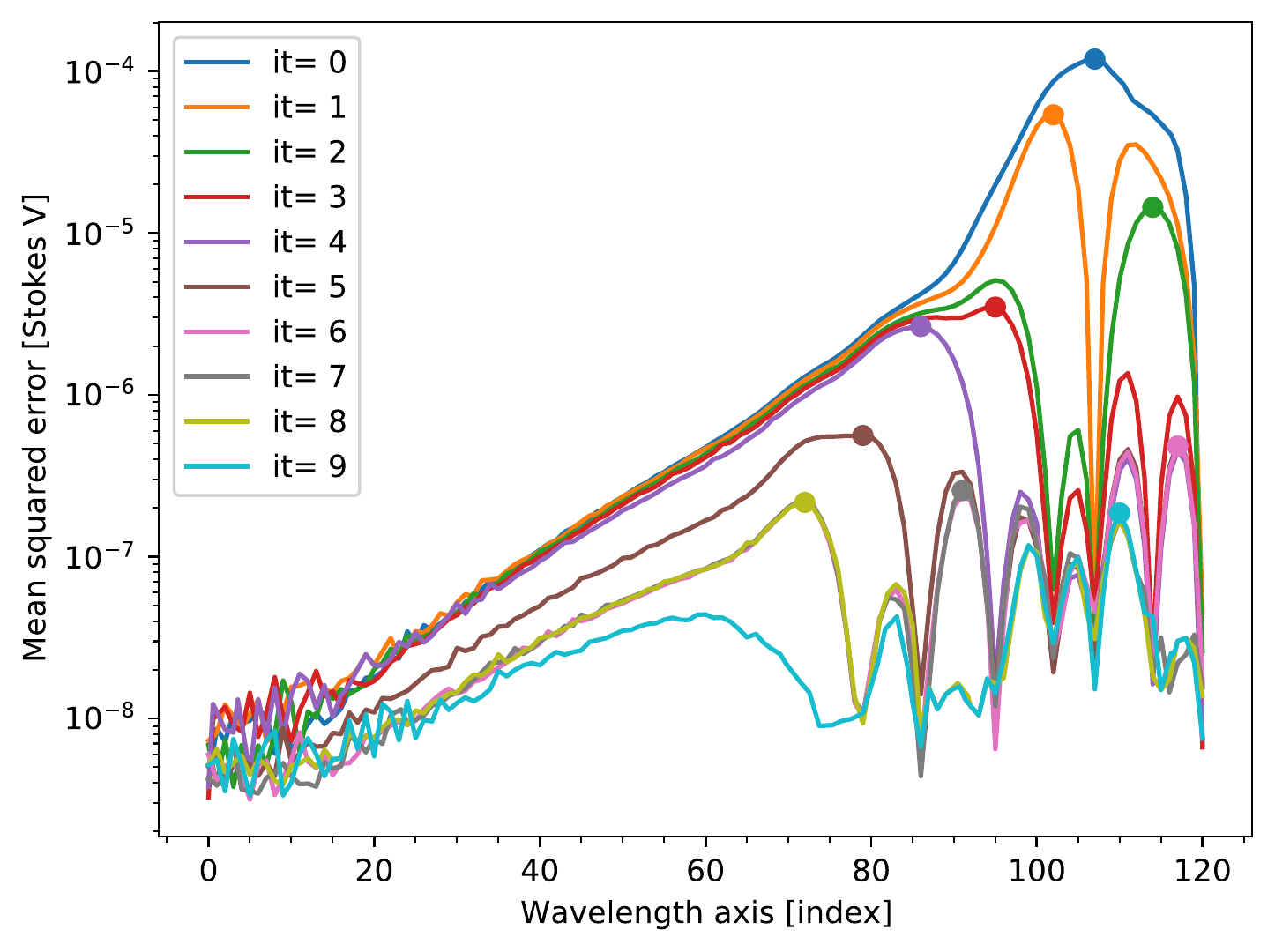}
\caption{Mean square error of the predictor at every iteration for Stokes $I$ and Stokes $V$ for the \ion{Ca}{ii} line at 8542\,\AA.} \label{fig:ca_error}
\end{figure}

\subsubsection{Complex example}

For a more complex test, we have chosen the chromospheric spectral line \ion{Ca}{ii} at 8542\,\AA. In this scenario, each profile can exhibit a wide range of shapes, including strong asymmetries and emission profiles. We have used the synthetic dataset of Stokes profiles and atmospheric models from \citealt{DiazBaso2022A&A} to have a realistic and diverse representation of observed profiles (quiet-sun, umbra, heating events, etc). They were obtained as the reconstructed atmospheric models from non-local thermodynamic equilibrium (NLTE) inversions using the STiC code \citep{delaCruz2016,delaCruz2019_STiC} on high-resolution spectropolarimetric data acquired with CRISP at the SST (more details in \citealt{DiazBaso2021}). This spectral line has a known asymmetry that arises from the presence of different isotopes \citep{Leenaarts2014}, and this new method could use this feature to its advantage. However, we will consider the spectral sampling symmetric to easily compare it with the symmetric sampling suggested by \cite{delaCruz2012A&A}, i.e., we will use half of the spectral line and the rest of the points should be mirrored. The dataset consists of 30000 Stokes profiles {with 240 points in the wavelength range $\pm$1.5\,$\AA$ and an equidistant sampling of 12.5\,m$\AA$}, all synthesized with an observing angle $\mu=1$. For a consistent comparison with the former sampling, we have assumed the same conditions: a Gaussian spectral filter transmission characterized by a full width at half maximum (FWHM) of 100 m$\AA$, an additive photon noise with a standard deviation $\sigma=10^{-3}$ relative to the continuum intensity, and the same number of points.

Once we have generated and convolved the Stokes profiles of the dataset, we follow the same procedure as before: we iteratively added the list of points with the largest errors according to the network prediction. Figure~\ref{fig:ca_sampling} shows the comparison with the sampling suggested by \citet{delaCruz2012A&A} in five different profiles. This sampling scheme and other similar versions have been used in previous SST observations \citep{Diaz2019_8542FIL, Yadav2021}. It comprises a fine grid of points in the core and a coarser grid in the wings that depends on the instrumental spectral FWHM at the wavelength of the line. It consists of points spaced 50\,m$\AA$ close to the core and every 500\,m$\AA$ in the wings. From the comparison, it is encouraging to see that our method provides a similar result compared to a strategy that has resulted from the analysis of the sensitivity of the line and its improvement over time.

Figure~\ref{fig:ca_error} shows the evolution of the average error in different iterations for Stokes $I$ and Stokes $V$. As a result, the final sampling scheme tends to have more points at those locations where the profiles show more variability, which are the knees (enhancement around $\pm0.2\AA$) and the core of the Stokes $I$ profiles, as well as the locations of the lobes of Stokes $V$. A similar result is found in the linear polarization parameters Stokes $Q$ and Stokes $U$. For Stokes $I$, we found a sampling scheme with an average distance of 70\,m$\AA$ near the core and over 450\,m$\AA$ near the wings. For Stokes~$V$, the average distance is about 65\,m$\AA$ in the core with no point until at the edge at $-1.5$\,$\AA$ because there is practically no signal in that region. These are results from the statistical analysis of the signals, but they already show the implications on the inference of physical quantities: observations focused on magnetic field analysis should prioritize a denser grid near the core, while those focused on thermodynamic properties benefit from a larger coverage \citep{Felipe2019A&A}. 

Although we have shown here how to implement this solution for individual Stokes parameters, it can also be possible to take all Stokes parameters at once into account in the same merit function and weigh them appropriately according to the desired requirements. If the predictor network is well trained (and the number of points is not too large), the location of the points also generally satisfies Nyquist's theorem and does not include points closer than half of the FWHM. This is expected because the spectral information has been degraded, so each spectral point contains information about the surrounding points and the network can use it to estimate the values of points close in wavelength. In fact, if the instrumental resolution is worse, the method automatically tends to increase the separation of the points.

We note that if our set of profiles includes large velocities, we will obtain a sampling scheme with a wavelength distance larger than the original spectral resolution of the instrument to capture the details of the core when it is shifted far from the rest wavelength. So the minimum distance in the sampling scheme depends on the spectral resolution of the instrument, the typical broadening of the spectral shape, and the expected Doppler velocities.

In summary, this strategy is very simple and only needs a pre-existing dataset with spectral examples of the line of interest. They can be generated from numerical simulations or by previous observations with an instrument with a better spectral resolution, such as slit-spectrographs.

\subsection{Using model information}\label{sec:model}
\subsubsection{Description of the method}

Although we have seen in Fig.\,\ref{fig:gauss2} that the model parameters are recovered better with the new sampling scheme, we have no control over which solar atmospheric variables are being improved the most. Therefore, if the physical model that generated that dataset is also available, we can optimize the spectral sampling to improve the inference of a given constraint or model parameter. {This is known as supervised selection as our target is now a product of the observed data.} This problem becomes much more general and non-linear than in the previous case where we were only interested in keeping as much information as possible. As it will be too expensive to train a network for each combination of wavelength points, we can go back to the sequential strategy of the previous approach to speed up the search. The main difference is that now the function to minimize is the mean square error with respect to the model parameters $T=\{\{t_q^k\}_{q=1}^Q\}^M_{m=1}$. Each sample $t^k$ is a $Q$-dimensional vector that can contain as many parameters as we want to infer. Thus, the pseudocode, in this case, would be:
\begin{algorithm}
	\caption{Future error-based spectral sampling} 
	\begin{algorithmic}[1]
	\State $L$ = pool of available points
	\State $U_0$ = set of initial points
	\State $T$ = parameters of interest
    \State $P$ = number of points to add to the scheme
		\For {$p=1,2,\ldots$ $P$}
		    \For {$x_m=1,2,\ldots N$}
    			\State \kern-0.5em train the predictor $f_\theta$ such that $T$ $\sim f_\theta$([$U_{p-1},(x_m,y_m)$])
    			\State \kern-0.5em evaluate the criteria $\sigma_m^2$ = mse($T$,$f_\theta$([$U_{p-1},(x_m,y_m)$]))
    	    \EndFor
			\State identify the most informative point: $i$ = argmin$(\{\sigma_m^2\})$
			\State add the new wavelength point $U_{p}=U_{p-1}+\{x_i,\{y_i^k\}_{k=1}^K\}$ 
		\EndFor
	\end{algorithmic} 
\end{algorithm}

\begin{figure}[ht!]
\centering
\includegraphics[width=0.95\linewidth]{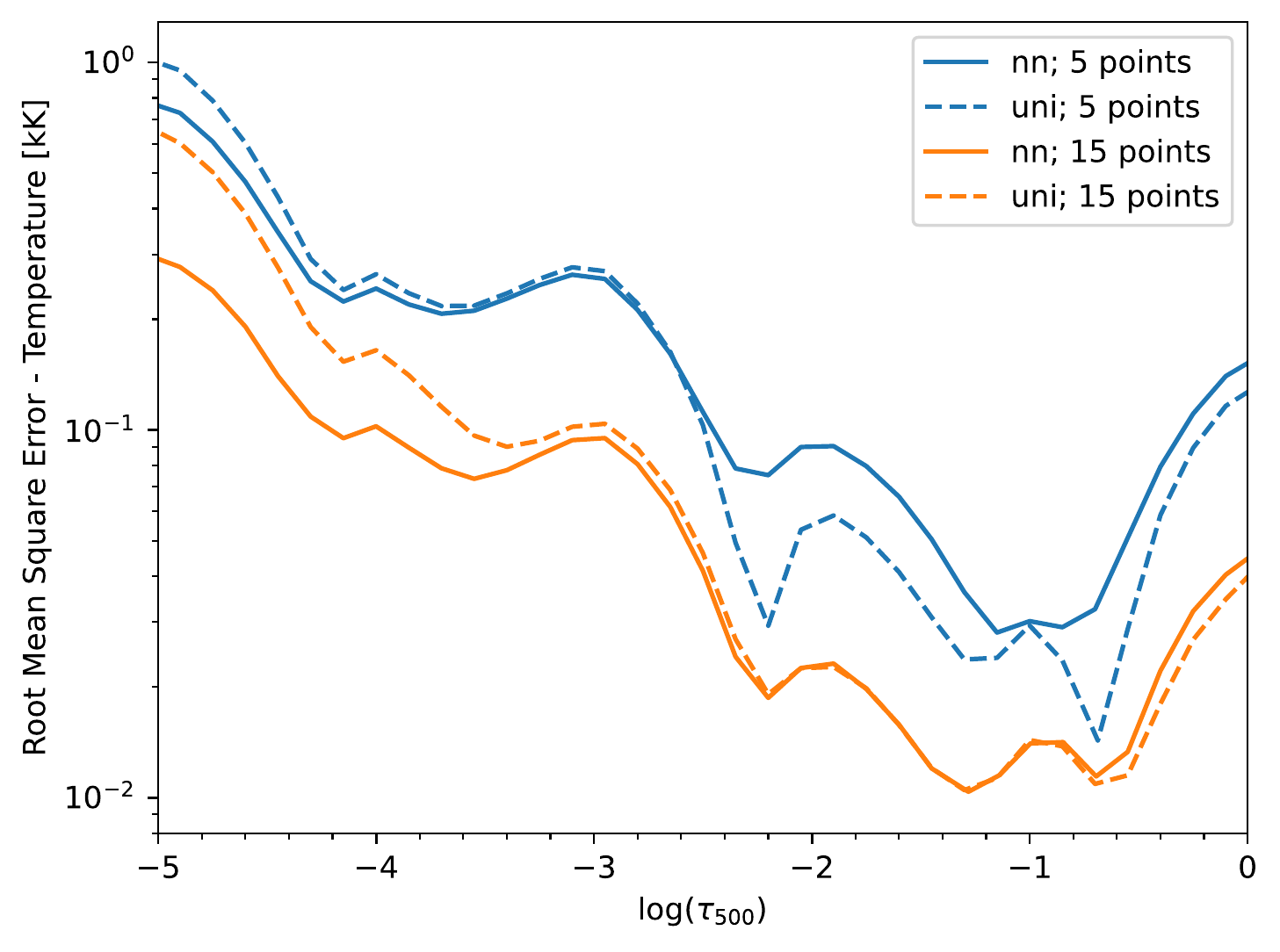}
\caption{Root mean square error of the network used as a predictor of the temperature stratification when having 5 points (blue lines) and when having 15 points (orange lines) in the whole wavelength range. The solid lines represent the error of the inferred sampling method and the dashed lines are produced using the uniform sampling scheme.} \label{fig:temp1}
\end{figure}

If we examine both approaches, we will see that while in the first one, we tried to minimize the error at the worst wavelength, in this method we will search for the new wavelength point that produces the best prediction. In this case, instead of training different neural networks for every configuration, the neural network has two inputs: one that is fixed (the sampling with the previous points) and a second input $(x_m,y_m)$ which is one of the points from the pool. During the training, we randomly sample points from the pool to train a single neural network that can predict the output no matter what point of the pool is chosen. To test this method, we have used the temperature stratification as a function of the optical depth scale at 5000\AA, hereafter $\log(\tau_{500})$, associated with the intensity profiles of the same dataset of the previous example. Figure~\ref{fig:temp1} represents the root mean square error according to the number of iterations of the temperature stratification.

As a result, the new scheme again tends to place more points in the core than in the wings, similar to the previous case. While in the last example this result was purely due to the variability of the spectral line, in this case, the reason (although related) is different. The two reasons are physically linked because the variability of the spectral line is due to the sensitivity of the line to changes in different regions of the solar atmosphere. To understand the correspondence in the physical parameters, we have also calculated the error for a uniform sampling scheme. Figure~\ref{fig:temp1} shows the error in the stratification of the temperature with 5 and 15 points in the full spectral range. From Fig.~\ref{fig:temp1} we see that the error-guided method tends to decrease the error in the chromosphere ($\log(\tau_{500})\simeq-4.5$) for a low number of points as the average error in the chromosphere is larger than in any other location. After having a few points in the core, the following iterations add points in the wings, which help to improve the estimation of the temperature of the lower atmosphere ($\log(\tau_{500})\simeq-1.5$). On the other hand, the uniform sampling is not guided by the uncertainty and follows a different behavior, improving first the lower atmosphere (for which we already had a good estimation) and later the rest of the atmosphere. This behavior occurs because the extension in wavelength where we have photospheric information is larger (few angstroms in the wings) compared with the narrow wavelength region that samples a large range of heights in the chromosphere \citep[see Fig.\,4 in][]{Kuridze2018ApJ}. In the end, we see that the remaining error cannot decrease any further since it does not depend on the number of wavelength points but the irreducible uncertainty resulting from the sensitivity of the line to the different heights in the solar atmosphere and the loss of information in the radiative transfer.

From this experiment, we show how a neural network can be used to indirectly estimate the average sensitivity of the profiles to the physical quantities of the model atmosphere, or in other words, the response functions \citep{RuizCobo1992, Centeno2022ApJ}. Given that the sensitivity of the observables to different parameters is also different, we also expect that optimizing the sampling to infer other quantities such as the line-of-sight velocity or the magnetic field will produce output schemes that will differ  from the other cases. As stated in the previous section, given that we are optimizing a single merit function, we could combine as many physical quantities as we want or weigh them across the stratification, so the technique is optimized for a particular region of heights in the solar atmosphere. 

We should recall that the training process of a neural network is non-deterministic and, the accuracy of the neural network is finite. This means we might arrive at a slightly different configuration after every run. This usually occurs with the last few points of the scheme, when the merit function has a very similar score everywhere. This problem could be alleviated with a network committee (ensemble), but in any case, the more points we add, the less essential it is to have an efficient design if we are already covering the whole area with many points. The use of the neural network as a surrogate function, in this case, allows us to overcome the very time-consuming process of inference in NLTE inversions. However, if a much simpler and deterministic model is available (e.g., the weak-field approximation) we recommend using such a model to eliminate the stochastic error inherent in the training process of the network, although simple but non-linear models (e.g., a Milne-Eddington model) also have their own stochastic problems in finding the solution depending on the optimization method used and parameter degeneracy.

We also note another caveat regarding the dataset we have used. In our case, we have synthesized the spectral lines assuming a that the mode atmosphere has not evolved during the time of the scan and only analyzed the impact of having different spectral resolution observations. A realistic study should include the scanning time effects when generating the sampling scheme, as shown by \cite{Felipe2019A&A} and \cite{Schlichenmaier2022}. This should help to quantify better the expected error in the inference when a specific temporal cadence is used.

\subsubsection{Model implications}\label{sec:implications}

\begin{figure}[ht!]
\centering
\includegraphics[width=0.9\linewidth,trim={0cm 0cm 0cm 0cm},clip]{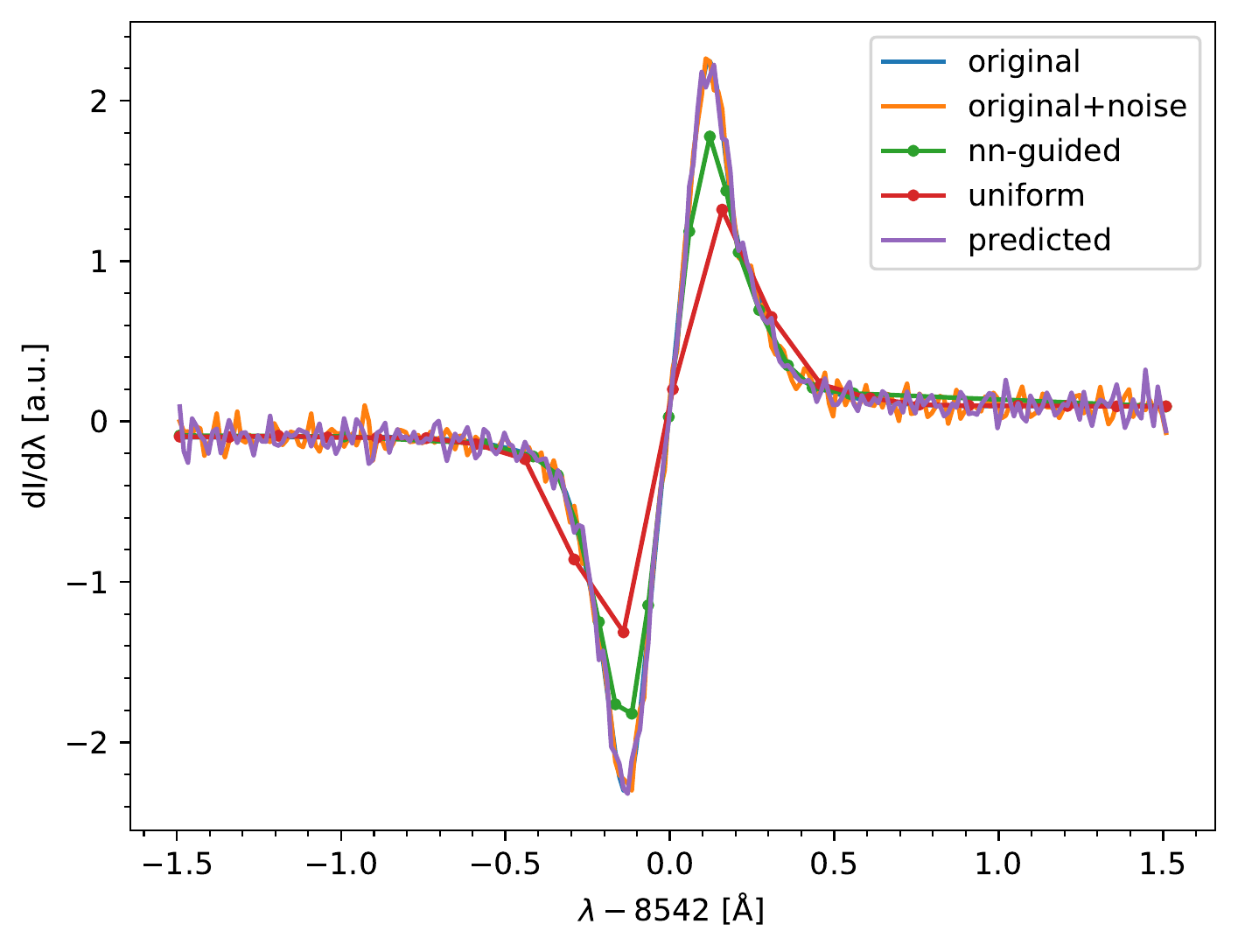}
\caption{Derivative of Stokes $I$ with respect to the wavelengths for the \ion{Ca}{ii} line at 8542\,\AA\ using the original synthetic profile, the sampling found in section~\ref{sec:spectra}, using a uniform sampling and using the prediction of the neural network for the full profile.} \label{fig:lasttest}
\end{figure}

In this last section, we want to show the implications of optimizing a wavelength scheme to the particularities of the chosen model. For this, we have chosen the weak-field approximation (WFA). In particular, we will optimize our scheme to calculate the line-of-sight component of the magnetic field following the maximum likelihood estimation of $B_{\rm LOS}$ in Gauss \citep{MartinezGonzalez2012MNRAS}:
\begin{equation}
    B_{\rm LOS}=\left(\sum_i V_i \cdot dI/d\lambda_i\right)/\left(C_0\cdot \lambda_0^2\cdot g_{\rm eff} \sum_i(dI/d\lambda_i)^2\right)
\end{equation}
where the constant $C_0=-4.67\cdot10^{-13}$ $\rm [G^{-1} \AA^{-1}]$, $\lambda_0$ in $\AA$ is the central wavelength of the line, and $g_{\rm eff}$ is the effective Landé factor. Since we do not have to train any neural network, the optimization becomes a straightforward calculation, where the derivative of the profile is calculated by the finite differences, using the centered derivatives formula for non-equidistant grids \citep{Sundqvist1970ASF}. Therefore, since Stokes~$V$ is the parameter that contains most of the information of $B_{\rm LOS}$, one would expect that the optimization would sample this quantity like the result of Fig.~\ref{fig:ca_sampling}. However, the estimation of the longitudinal field also depends on the derivative of Stokes~$I$, and a poor sampling of Stokes~$I$ will also have a strong impact. In fact, the output optimized scheme has the location of the points extremely close to each other so that the truncation error in the derivative calculated with finite differences is smaller than the uncertainty introduced by Stokes~$V$. If we provide the method with the exact value of the derivative, the method then tends to accommodate more points at the location of the Stokes $V$ lobes. Given this result, optimizing the sampling to increase the accuracy to $B_{\rm LOS}$ does not seem to be the smartest choice if we want to extract other information. A better idea would be to use the scheme in Fig.~\ref{fig:ca_sampling} and use the trained network to interpolate the Stokes~$I$ profile to a finer mesh where we can compute the derivative. In Fig.~\ref{fig:lasttest} we show the effect on the derivative of the Stokes~$I$ profile for a given example using various methods. In general, the derivative of the profile estimated with the neural network has higher accuracy, although other spline-based interpolation methods could be used when having a large number of points. For a low number of points, the neural network performs much better. In terms of magnetic fields, for the case of a scheme with 21 points and a noise level of $10^{-3}$ in units of the continuum, the root mean square error for this database is of the order of 800\,G for the uniform case, 200\,G when using the sampling guided by the neural network found in section~\ref{sec:spectra}, 60\,G when the same neural network is used for predicting the full profile and 30\,G for the case of of having a very fine grid in Stokes $I$ and $V$.

Therefore, we conclude that (for FPIs observations) using models that fit the Stokes profiles directly (such as a Milne-Eddington model) could be more accurate than models, like the WFA, that depends also on the derivatives of the Stokes profiles. Recently, a modified Milne-Eddington approximation has been developed to allow much better modeling of chromospheric lines \citep{Dorantes2022A&A} if speed is needed. On the other hand, if precision is required, we recommend NLTE modeling for the inference.

\subsection{Sampling parameterization}
\begin{figure}[ht!]
\includegraphics[width=\linewidth,right,trim={0cm 1.2cm 13cm 0cm},clip]{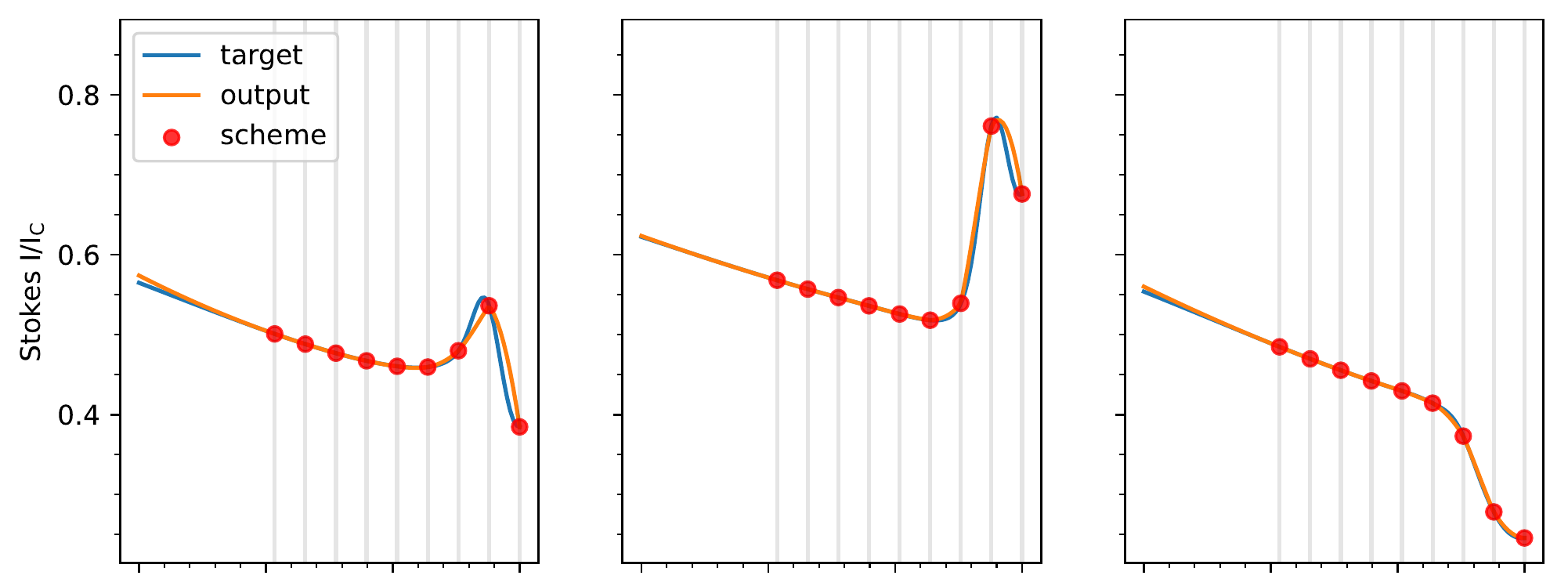}
\includegraphics[width=\linewidth,right,trim={0cm 0.3cm 13cm 0cm},clip]{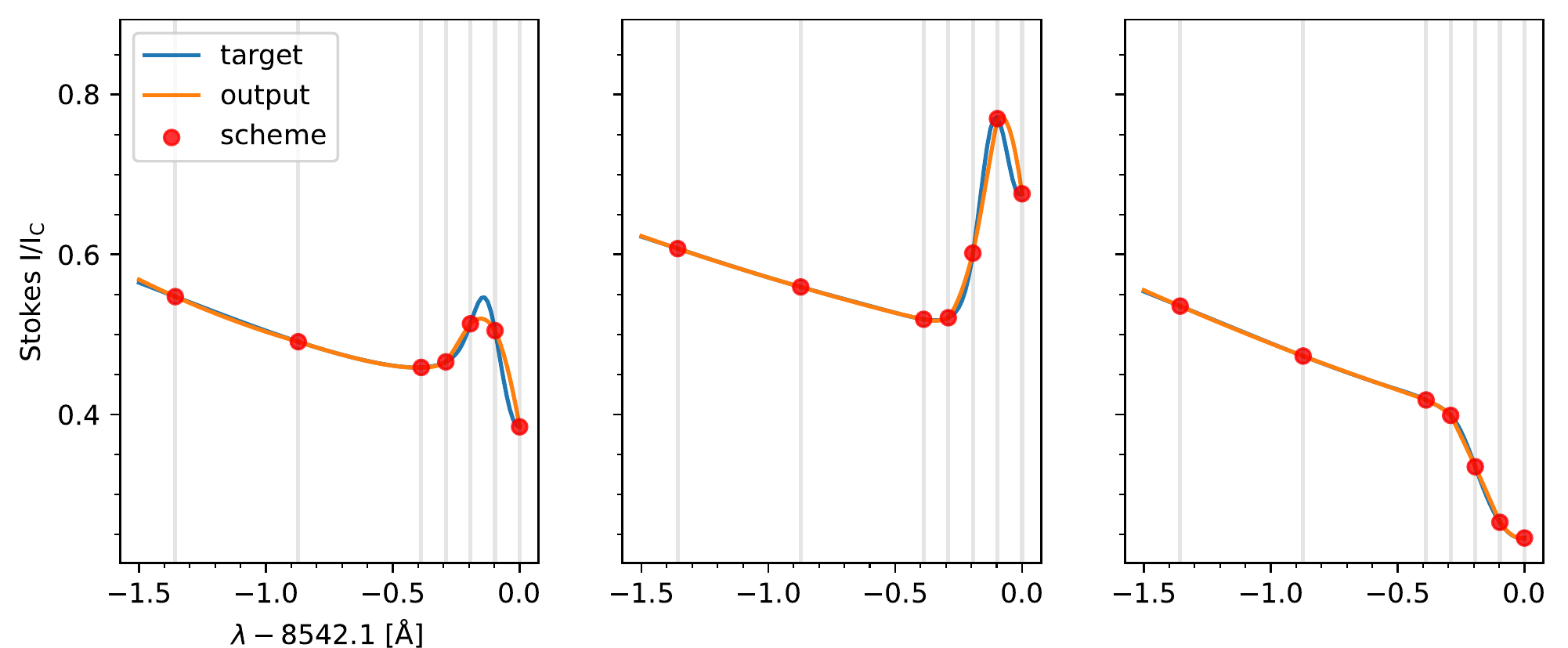}
\caption{Optimal sampling where all points have the same distance (upper row) and where outer points are spaced multiple times wider than the inner points (bottom row).} 
\label{fig:parametrized1}
\end{figure}

{Finally, although we have shown a method flexible enough to find the points in wavelength that retain the most information and improve our parameter inference, in some cases we may want the distances of the points to obey other constraints, such as being multiples of a uniform grid to perform computationally efficient fast Fourier transforms, because of instrumental reasons, or just because the interface of some spectral inversion codes requires that. In this case, one can parameterize the sampling (for example, by defining the inner distance, distance factor for outer points, and the number of inner points) and train our predictor for several cases. Later we can evaluate the different configurations (in this three-dimensional space) and find which configuration is optimal. An example of this idea is shown in Fig.~\ref{fig:parametrized1} where we show the optimal configuration for a uniform sampling (top row) and another in which we allow the outer points to be further apart (bottom row). As a result, we find a 13-point scheme (step of 95\,m$\AA$, 7 inner points, an outer factor of 5) whose performance is better than a completely uniform scheme with 19 points (step of 120 m$\AA$). Both are the best configurations for their respective number of points to have a fair comparison. Using this strategy we can find schemes with fewer points while obeying our constraints.}

\section{Conclusions}\label{sec:conclusions}

In this study, we have explored, developed, and validated a machine-learning technique to design efficient sampling schemes for accurate parameter estimation using filtergraph instruments. {The problem of finding an optimal wavelength configuration has been framed as a feature selection problem.} Only those points (features) that contain relevant information and improve the model prediction are chosen to be in the sampling scheme. {We have implemented two approaches: if only the spectra is available (unsupervised selection) or if the model information is used to improve the inference of the physical quantities (supervised selection). We have implemented a sequential approach where a neural network is used to predict the target from the information at few points, and in that way we can measure the importance of different wavelength points. We have validated the performance of the new sampling schemes compared to a uniformly spaced sampling by showing the lower errors in the reconstructions of the model parameters. This approach is particularly efficient for a small number of points, which is crucial when observers are interested in highly dynamic events and a good temporal cadence is needed.}
Also if we expect large Doppler velocities or the spectral resolution of our instrument is poor, this method will adapt the scheme accordingly. {Both the method description and the code implementation are publicly available}.

From the experiments using the \ion{Ca}{ii} 8542\,\AA\ line we could extract some results that can be used as practical recommendations for future observations. For example, the resulting wavelength scheme naturally places (almost a factor 4) more points in the core ($\pm 0.5\,\AA$) than in the wings, consistent with the sensitivity of the spectral line in each wavelength range. In addition, observations focused on magnetic field analysis should prioritize a denser grid near the core {where the polarization signals are present}, while those focused on thermodynamic properties benefit from a larger coverage \citep{Felipe2019A&A}. In conclusion, the selection of the wavelength points depends on the goal of the observations.

Regarding the methodology, it is difficult to ensure that this technique provides the global solution to the problem (if there is such), but it provides an efficient scheme, much better than a uniform strategy. Therefore, this method can be helpful when designing new instrumentation or when designing the wavelength configuration in observing proposals for a specific target. This trained network can also be used after the observations to estimate the complete profile from the points obtained, thus helping the inference process when using NLTE inversion codes \citep{delaCruz2019_STiC} or by {obtaining} a more accurate estimation of the derivative of Stokes\,$I$ when using the weak-field approximation for inferring the magnetic field.

Finally, there are several ways in which we could improve the current implementation. For example, by solving the full problem at once because the neural network makes the problem differentiable. Using gradient-based techniques we could find the solution by applying a regularization term (e.g. $\ell_0$ norm) that penalizes a large number of points. We have tried this option but it was not possible to converge in most cases as the full problem is much more degenerated (i.e., having the location of all wavelength points as free parameters at the same time) than using a sequential way. Regarding the ultimate goal of the technique, this procedure is so flexible that it could also be generalized to include the exposure time at each wavelength. By doing so, the merit function would be sensitive to the signal-to-noise ratio at each point, and we could obtain a better design by taking into account the trade-off between the weak polarization signals and the evolution time of the Sun. From the neural network perspective, there are also ways to speed up the training and decrease the stochasticity of the process by using networks that better exploit the relation between different regions of the spectrum (such as Transformers or Graphs networks) or by reformulating the problem with reinforcement learning to mitigate the nested behavior of sequential approaches. In summary, we plan to continue investigating this strategy and provide optimal wavelength schemes for different spectral lines, instrumentation, and solar targets in future work.

\begin{acknowledgements}

{We would like to thank the anonymous referee for their comments and suggestions.}
CJDB thanks Ignasi J. Soler Poquet, João M. da Silva Santos and Henrik Eklund for their comments.
This research is supported by the Research Council of Norway, project number 325491 and through its Centers of Excellence scheme, project number 262622.
This project has received funding from the European Research Council (ERC) under the European Union's Horizon 2020 research and innovation program (SUNMAG, grant agreement 759548).
%
The Institute for Solar Physics is supported by a grant for research infrastructures of national importance from the Swedish Research Council (registration number 2021-00169).
We acknowledge the community effort devoted to the development of the following open-source packages that were used in this work: NumPy (\url{numpy.org}), Matplotlib (\url{matplotlib.org}), SciPy (\url{scipy.org}), Astropy (\url{astropy.org}) and PyTorch (\url{pytorch.org}).
This research has made use of NASA's Astrophysics Data System Bibliographic Services.
\end{acknowledgements}


\bibliographystyle{aa}

\end{document}